\documentclass[two column,jgrga]{AGUTeX}
\usepackage[dvips]{graphicx}
\graphicspath{{figure1/}}
\usepackage{amssymb}
\authorrunninghead{Savani ET AL.}

\titlerunninghead{momentum flux of CMEs}

\authoraddr{Corresponding author: N. P. Savani,
Solar Physics Branch, Naval Research Laboratory, 4555 Overlook Ave SW,
Washington, DC 20375, USA.
(neel.savani02@imperial.ac.uk)}

\begin{document}

%% ------------------------------------------------------------------------ %%
%
%  TITLE
%
%% ------------------------------------------------------------------------ %%

\title{Tracking the momentum flux of a CME and quantifying its influence on geomagnetically induced currents at Earth}
%
% e.g., \title{Terrestrial ring current:
% Origin, formation, and decay $\alpha\beta\Gamma\Delta$}
%

%% ------------------------------------------------------------------------ %%
%
%  AUTHORS AND AFFILIATIONS
%
%% ------------------------------------------------------------------------ %%

%Use \author{\altaffilmark{}} and \altaffiltext{}

% \altaffilmark will produce footnote;
% matching \altaffiltext will appear at bottom of page.
\authors{N. P. Savani,\altaffilmark{1,2} A. Vourlidas,\altaffilmark{3} A. Pulkkinen,\altaffilmark{2,4} T. Nieves-Chinchilla,\altaffilmark{2,4} B. Lavraud,\altaffilmark{5,6}  M. J. Owens,\altaffilmark{7}}

% R. Kataoka,\altaffilmark{5}

% N. P. Savani,\altaffilmark{1,6} A. Vourlidas,\altaffilmark{2} A. Pullkinen,\altaffilmark{3,6} N.R. Sheeley,\altaffilmark{2}, L. Simpson, \altaffilmark{2} T. Nieves-Chinchilla,\altaffilmark{3,6} R. A. Howard,\altaffilmark{2} B. Lavraud,\altaffilmark{4} R. Kataoka,\altaffilmark{5} A. P. Rouillard,\altaffilmark{4}

\altaffiltext{1}{University Corporation for Atmospheric Research (UCAR), Boulder, Co, USA }

\altaffiltext{2}{NASA Goddard space flight center, Maryland, USA}

\altaffiltext{3}{Space Science Division, Naval Research Laboratory, Washington, DC, USA}

\altaffiltext{4}{Catholic University of America, Washington, DC, USA.}

\altaffiltext{5}{Institut de Recherche en Astrophysique et Planétologie, Université de Toulouse (UPS), France}

\altaffiltext{6}{UMR 5277, Centre National de la Recherche Scientifique, Toulouse,
France}

\altaffiltext{7}{Space Environment Physics Group, University of Reading, RG6 6BB, UK}
%\altaffiltext{5}{Tokyo Tech, Tokyo, Japan}

%% ------------------------------------------------------------------------ %%
%
%  ABSTRACT
%
%% ------------------------------------------------------------------------ %%

% >> Do NOT include any \begin...\end commands within
% >> the body of the abstract.
% We use a forward modelling technique to estimate the shape and orientation of the CME and use it to implement a semi-automated method for estimating the mass of the CME until ~30Rs.

\begin{abstract}
We investigate a CME propagating towards Earth on 29 March 2011. This event is specifically chosen for its predominately northward directed magnetic field, so that the influence from the momentum flux onto Earth can be isolated. We focus our study on understanding how a small Earth-directed segment propagates. Mass images are created from the white-light cameras onboard STEREO which are also converted into mass height-time maps (mass J-maps). The mass tracks on these J-maps correspond to the sheath region between the CME and its associated shock front as detected by in situ measurements at L1. A time-series of mass measurements from the STEREO COR-2A instrument are made along the Earth propagation direction. Qualitatively, this mass time-series shows a remarkable resemblance to the L1 in situ density series. The in situ measurements are used as inputs into a 3D magnetospheric space weather simulation from CCMC. These simulations display a sudden compression of the magnetosphere from the large momentum flux at the leading edge of the CME and predictions are made for the time-derivative of the magnetic field (dB/dt) on the ground. The predicted dB/dt were then compared with observations from specific equatorially-located ground stations and show notable similarity. This study of the momentum of a CME from the Sun down to its influence on magnetic ground stations on Earth is presented as preliminary proof of concept, such that future attempts may try to use remote sensing to create density and velocity time-series as inputs to magnetospheric simulations.
\end{abstract}

%% ------------------------------------------------------------------------ %%
%
%  BEGIN ARTICLE
%
%% ------------------------------------------------------------------------ %%

% The body of the article must start with a \begin{article} command
%
% \end{article} must follow the references section, before the figures
%  and tables.

\begin{article}

%% ------------------------------------------------------------------------ %%
%
%  TEXT
%
%% ------------------------------------------------------------------------ %%

\section{Introduction}

The influence of a southward orientated magnetic field of the solar wind impinging onto the Earth’s magnetosphere is known to be the main driver for coupling the solar wind to the terrestrial system \citep[e.g.][]{gonzalez1974, bargatze1985, tsurutani1992}. In this case, the oppositely oriented magnetic fields between the solar wind and the magnetosphere are more easily able to reconnect and thereby transfer energy, mass and momentum to the Earth. For this reason significant emphasis on understanding and predicting the magnetic field orientation has been pursued to improve our space weather predictive capabilities.

The role of the solar wind density, $N_{sw}$, in space weather studies is more complex. Statistical studies aimed at coupling the solar wind density to geomagnetic indices such as Dst show a weak correlation \citep{obrien2000}. Others suggest that $N_{sw}$ should be significant in mediating the energy transferred to the magnetosphere \citep{borovsky1998, thomsen1998}. The density has been demonstrated to affect the magnetospheric response by way of the solar wind dynamic pressure \citep[momentum flux, e.g.][]{xie2008}. A sharp increase in solar wind dynamic pressure can rapidly compress the Earth's magnetopause causing a large time rate of change in the local ground-based magnetic field intensity. While this effect is not often considered as a measure of `geo-effective' events, the space weather consequences can be severe. An example of this can be found in the 4th August 1972 event, where the peak Dst index was moderate compared to resulting outage of the AT$\&$T telecommunications cable and the geosynchronous satellite solar cell damage \citep{anderson1974, lanzerotti1992, tsurutani1992b} .

%used impulse response filtering to investigate the geoefficiency of the response of the Dst to the solar wind motional electric field (vB). The $N_{sw}$ and the momentum flux was shown to have similar impulse responses, while the velocity showed little dependency. Therefore Weigel concluded that the response in Dst which was used as a proxy to the ring current is highly dependent on $N_{sw}$.

When investigating the drivers of space weather at Earth and in particular geomagnetically induced currents (GICs), it is important to distinguish between the different solar wind structures that initiate the chain of events that may lead to significant socio-economic losses. Different interplanetary structures have been previously reported to produce significantly varying responses in the Earth magnetosphere \citep[e.g.][]{huttunen2008, miyoshi2005, borovsky2006, denton2006, lavraud2008}, it is therefore reasonable to expect different responses for GICs. The two main drivers of GICs are from intense magnetic storms associated with interplanetary coronal mass ejections (CMEs) and co-rotating interaction regions (CIRs). \citet{kataoka2008}  and \citet{borovsky2006} both found that the biggest problems for ground-based conducting systems were from CME drivers, whereas CIR-driven storms had a relatively minor effect.

Fast moving interplanetary CMEs can be considered to be made of two primary constituents: the sheath region between the shock front and the leading edge of the CME; and the ejecta itself. The ejecta is often magnetically dominated (i.e. plasma  $\beta<$1) and displays the magnetic properties of a flux rope (FR) when seen in situ \citep{burlaga1988}. The sheath region however contains denser compressed plasma with a higher value of plasma $\beta$. The direction of the magnetic fields within this region is variable, but is often found to vary within a 2D plane \citep{nakagawa1989, jones2002, kataoka2005, savani2011b}. The momentum flux, $P_{dyn}$= $N_{sw}V^{2}$ within the sheath is typically high. \citet{huttunen2008} concluded that as a geo-effective CME propagates over the Earth (usually for a period of $\sim24$ hours), the most intense GIC activity recorded on the ground is most likely to occur at the beginning, during the passage of the sheath region, between the shock and the leading edge of the magnetic obstacle.

A high solar wind density can cause a change in the compression ratio of the bow shock, as is frequent within a geomagnetic storm \citep{lopez2004}. For southward magnetic field this serves to increase the sensitivity of the predicted effects to $N_{sw}$ \citep[see also][] {lavraud2008}. Recent studies have considered the effect of preconditioning the magnetosphere to increase the space weather effects. For geomagnetic storms driven by both CMEs and CIRs, \citet{lavraud2006} showed that the Dst were under-predicted by a model for intervals that were preceded by an extended period of northward IMF. This is consistent with the hypothesis that a preconditioned high density plasma sheet will cause a larger than expected geomagnetic storm, as also expected from simulations \citep{lavraud2007}. The density of the Earth's plasma sheet is regulated by the solar wind density and lags behind by a few hours \citep{borovsky1998}.  For CME-driven storms the plasma sheet is generally more dense and persists for longer than those driven by CIRs \citep{denton2006}. A magnetohydrodynamic (MHD) simulation studied by \citet{siscoe2002} found that the polar cap potential saturation follows a power law with $P_{dyn}$. This then led \citet{xie2008} to develop a model where the magnetosphere can be preconditioned by $P_{dyn}$.

The Dst index is often used as a measure of the severity of a geomagnetic storm. This index is an average estimate of the global response as determined by four equatorially located ground stations. The Dst measures the horizontal component of Earth's ground magnetic field and is predominately monitoring the strength of the ring current in the magnetosphere \citep{fukushima1973,liemohn2001}. Since a denser and cooler plasma sheet produces a stronger ring current, CME-driven storms often produce more intense Dst values. Also, low latitude ($<56^{\circ}$) aurora, as measured in Japan were found to be predominately CME-associated \citep{shiokawa2005}. During a CME-triggered geomagnetic storm, the Dst profile often begins with a short duration rise which is triggered by the arrival of the CME-shock front (sudden storm commencement, SSC). The duration of the rise phase between the SSC and storm onset (SO) often corresponds to the sheath region upstream of the magnetic ejecta of the CME. The main phase of the geomagnetic storm, which corresponds to the magnetic ejecta traversing over the magnetosphere, is where a large decrease in the Dst value occurs. However, the phases of a geomagnetic storm can be time shifted under conditions where a strong southward Bz persists within the sheath region. The amplitude of the Dst decrease during the main phase is closely related to the southward magnetic field in the IMF and its subsequent magnetic reconnection with the magnetosphere.

The momentum flux of a CME, and in particular the sheath region, is higher than that of the ambient solar wind. This introduces a changing pressure onto the magnetosphere, and in the case of a CME-driven shock front, the change is abrupt.  The subsequent compression of the magnetosphere is expected to generate currents within the terrestrial system which are then detected on the ground as magnetic field fluctuations. As displayed by a Dst time profile during a geomagnetic storm, the geomagnetic field undergoes fast changes over a short time period. The largest time derivative of the ground magnetic field, dB/dt, occurs during the initial phase (between SSC and SO) or during substorms. As indicated by Faraday's law of induction, dB/dt is key in estimating the geomagnetically induced currents in technological conductor systems like power grids \citep{pulkkinen2006, baker2008}. Therefore dB/dt is often considered as a reasonable proxy for localised GIC activity \citep{viljanen2001}. In order to calculate the geoelectric field, additional information on the ground conductivity and specifics on the technological systems involved are required \citep{pirjola2002, pulkkinen2007}. In this paper we will focus on tracking the effects of CME momentum flux to the proxy of GICs, namely dB/dt.

Significant geomagnetically induced currents have been observed to affect technological systems in mid- and low-latitude region \citep[e.g.][]{ngwira2008, liu2009, watari2009}; and therefore it is becoming increasingly clear that these geomagnetically storm-driven GICs are not just a high latitude phenomenon \citep{pulkkinen2010}. The extent to which GICs can be expressed by dB/dt is also not as straightforward as first expected due to the complications of estimating the ground conductivity and how this varies at different depths. \citet{pulkkinen2010} showed that in Japan where the tectonic plates are geologically active, the local subduction zone is able to affect GICs in an unusual way by following the amplitude of the local geomagnetic field rather than the time derivative; this highlights the complexity of predicting GICs.

In this paper we focus on better characterising the $ N_{sw}$ from remote observations as a solar wind driver for dB/dt of the geomagnetic field. We take steps towards creating a time series of the solar wind density, and show that the momentum flux of a CME is an important parameter when predicting space weather incidents. We show this by comparing the ground station data with predicted estimates from the space weather modeling framework (SWMF) \citep{toth2005}.
% flux of a CME is an important parameter in mitigating against damaging space weather incidents by comparing the ground station data with predicted estimates from the space weather modeling framework (SWMF) \citep{toth2005}.

\section{Remote-sensing observations}
\begin{figure}
\noindent\includegraphics[width=23pc]{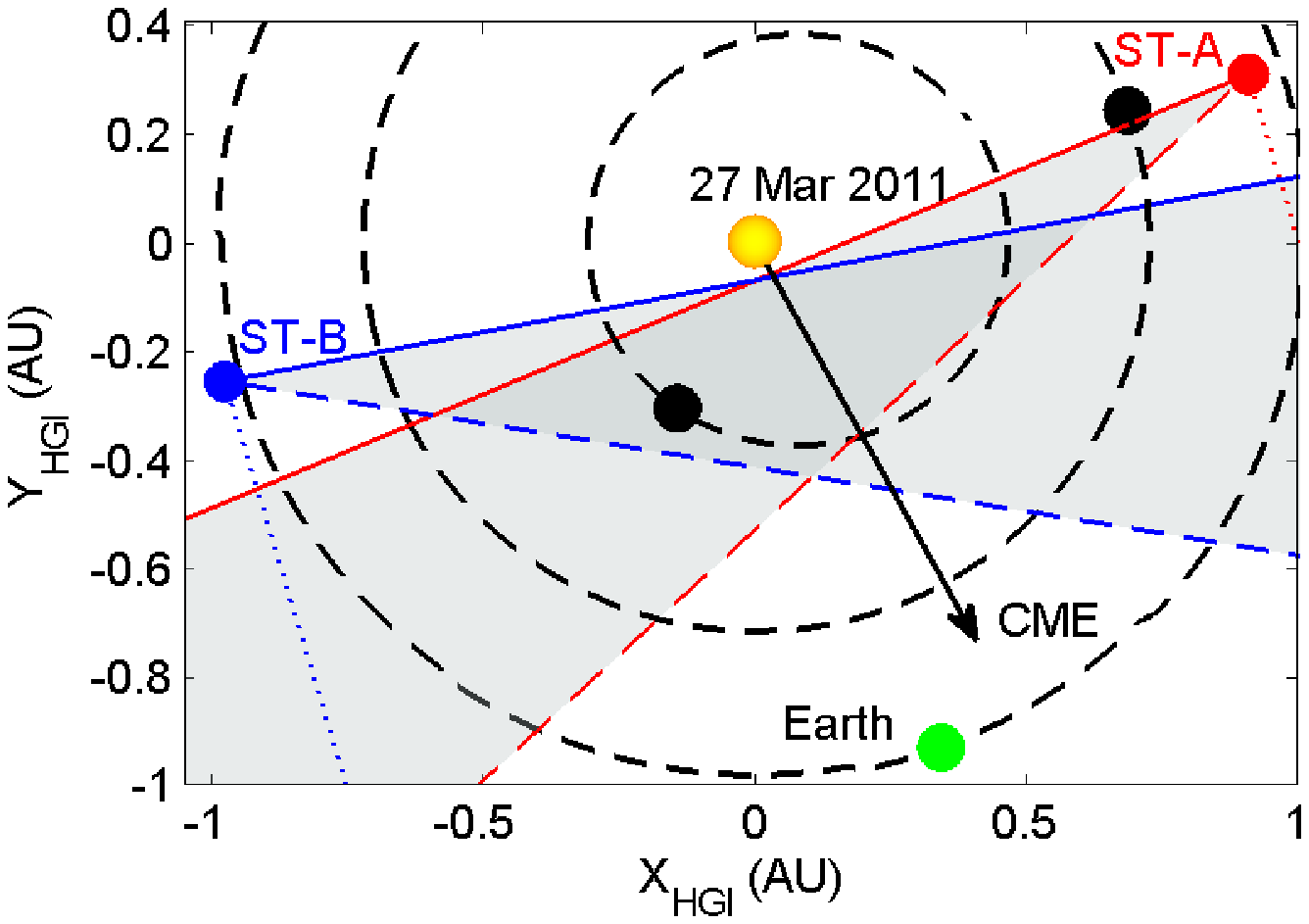}
\caption{Location of spacecraft during CME propagation on the 27 March 2011. The shaded regions locate the field of view for the HI-1 cameras on both the ST-A and ST-B spacecraft.}
\label{scPos}
\end{figure}

The STEREO mission, launched in 2006 \citep{kaiser2008}, consists of two spacecraft that follow a trajectory similar to that of the Earth. As they separate from each other at a rate of $\approx 45^{\circ}$ per year, one spacecraft travels ahead of the Earth (ST-A) while the other lags behind (ST-B). Each spacecraft carries the Sun Earth Connection Coronal and Heliospheric Investigation \citep[SECCHI,]{howard2008} imaging package, which contains an Extreme Ultraviolet Imager (EUVI), two coronagraphs (COR-1 and COR-2), and the Heliospheric Imager (HI). The HI instrument on each STEREO spacecraft is made up of two wide-field visible-light imagers, HI-1 and HI-2 \citep{eyles2009}. The fields of view of HI-1 and HI-2 are of $20^{\circ}$ and $70^{\circ}$ angular extent, respectively, and under ordinary operation are nominally centred at $13.7^{\circ}$ and $53.4^{\circ}$ elongation in the ecliptic plane. Thus the ecliptic plane corresponds to a horizontal line that runs through the centre of the fields of view. Figure \ref{scPos} displays the location of the STEREO spacecraft in relation to the Sun and Earth on 27th March 2011. The shaded regions indicate the field of view for the HI-1 cameras on both spacecraft. The direction of propagation for the CME nose is estimated from Jmap techniques \citep[e.g.][]{sheeley1999, rouillard2008, davies2009, savani2009, savani2012a} and detailed below in \S ~\ref{mjmap}.

	\subsection{CME mass calculation}
In order to convert the white light images that contain photometric information per pixel into mass per pixel we exploit the Thomson scattering properties of electrons from the corona and inner heliosphere. The total light observed by the coronagraphic and heliospheric imagers onboard STEREO are from photospheric photons scattered by all the electrons along the entire line of sight (LOS). The difference in contributions of each electron along the LOS depends on the distance and the Thomson scattering mechanism \citep{billings1966, vourlidas2006}, therefore an estimate of the total number can be generated from white-light images \citep{hayes2001}. In order to estimate the mass of a CME, the brightness contribution of the transient must be isolated from the background coronal signal. Previous studies have achieved this by subtracting a suitably-chosen pre-event image from an image containing the CME \citep{stewart1974, howard1985, poland1981}. A comprehensive explanation of the procedures required to make estimates of a CME mass from the raw telemetry data received on the ground is given by \citet{vourlidas2010}. However for completeness, the main key points in the procedure are given below:

1. First, the relevant time stamps for images containing the CME are chosen as well as a pre-event image. The pre-event image is ideally the last possible image prior to the CME entering the field of view in order to make the appropriate corrections for background. It is important to minimise the effects from evolutionary changes and solar rotation, and not to include another CME or other transient effects in the pre-event.

2. The pre-event image is subtracted from the sequence of CME-event images. These calibrated images now display information on the excess (depletion) of light in units of mean solar brightness, MSB (i.e a base-difference sequence).

3. The excess (depletion) of MSB within each pixel in the image is converted into a number of electrons by using the Thomson scattering equations and by assuming all the electrons are located on a single plane determined by the fixed-phi J-map technique.

4. The mass per pixel is calculated from each image by assuming a solar wind distribution of 90\% H and 10\% He. This corresponds to a mass of $1.97 \times 10^{-24}g $ per electron \citep{hildner1975}.

5. The mass of the CME is then estimated by summing up the values of all the pixels containing the CME.

\begin{figure}
\noindent\includegraphics[width=44pc]{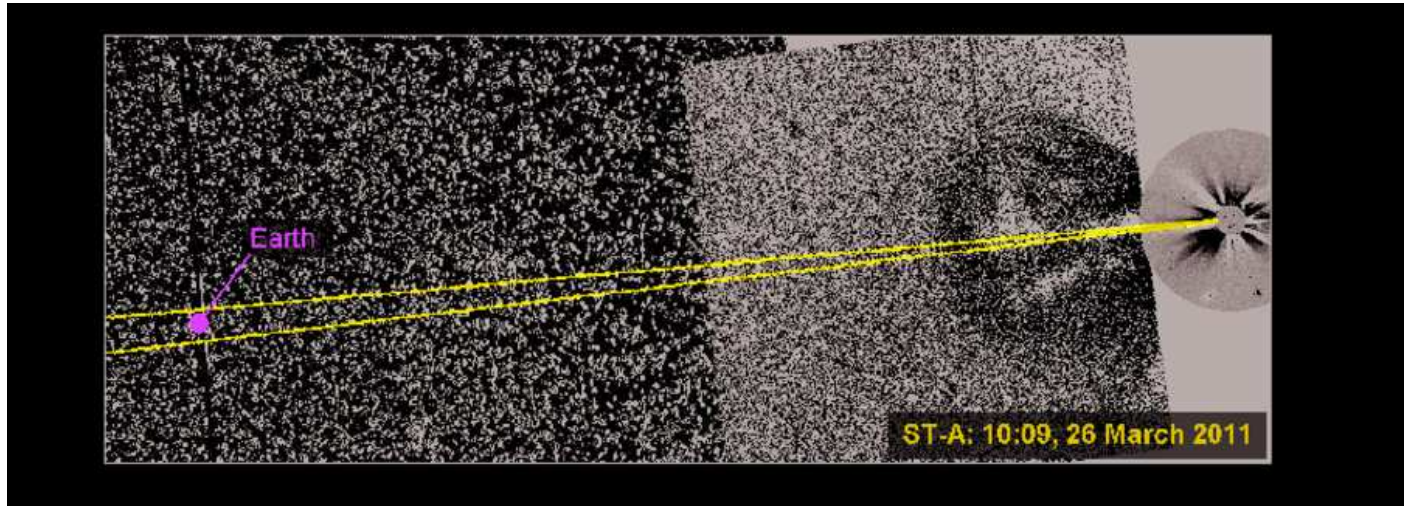}
\caption{A mass image from the COR2, HI-1 and HI-2 cameras from ST-A. The pixel intensity displays the locations of highest line of sight mass measurements. The cone overplotted on the figure displays the relevant part of the CME that propagates over the Earth.}
\label{HIcam}
\end{figure}

Figure \ref{HIcam} is a combined and cropped image from the ST-A and includes a frame from COR-2, HI-1 and HI-2. Practically speaking, the pre-event image is usually subtracted from the frames which occur later in time (i.e. when the CME has entered the field of view). However, this frame may also be subtracted from images that occurred earlier. In figure \ref{HIcam}, the image from the HI-2A camera is one that is actually earlier in time than the pre-event image. Therefore we are demonstrating the feasibility of using the pre-event image as a static background for frames earlier and later in time.
%As shown in figure \ref{HIcam}, the image from the HI-2A camera is one that is actually earlier in time than the pre-event image because the CME is within HI-1A. Therefore we are demonstrating the feasibility of subtracting the pre-event image from other frames that are prior to the CME entering the frame. Figure \ref{HIcam} is a combined and cropped image from the ST-A and includes a frame from COR-2, HI-1 and HI-2. 

The pre-event image subtraction process is most suitable for relatively short time-scales due to the steady state assumption. This assumption therefore begins to breakdown for the long duration of the CME within the large field of view. Also, HI-2 is sensitive to the background star field. This means that the ability to detect a CME motion is reduced if the star field is not appropriately removed during the initial image processing prior to making mass calculations. This paper is focusing on the potential to track the momentum flux and to monitor the relevant effects at Earth. For this reason we choose to use the simplest image processing at this stage in order to emphasise the minimum capability for space weather forecasting. Further processing of the images should be able to improve the tracking of the CME momentum \citep{howardt2012, howardt2012b} and may provide a more sensitive time series of mass flux at the L1 point.

The majority of the current work on mass estimates use data from coronagraphs which have a small field of view and assume the Thomson sphere is a flat plane. For the large field of view for both the HI cameras this is inappropriate. When estimating the mass, the propagation direction, and therefore the angle away from the plane of sky (PoS), is important in calculating the amount of Thomson scattering from electrons; i.e. the pixel on the inner edge of the camera has a different angle away from the PoS to the outer edge. This modification to the calculations is made in our work. In this paper, we treat each pixel individually as a different angle away from the plane of sky as measured along the line of sight.

	\subsection{Semi-automated CME mass}
The calculation of the total CME mass is carried out by summing up the mass values of each pixel within the observed CME. This region can be defined in a variety of ways: 

1. The sector method \citep{vourlidas2010}. The observer manually defines a set of four boundaries that can be used for the entire sequence of images. These boundaries are defined between two Position angles (P.A.) and the inner and outer radial boundary.

2. Region of interest (ROI) method \citep{vourlidas2000, subramanian2007}. The observer manually draws an outline of the CME. All the pixels within this ROI is considered to contribute towards the CME mass. This method calculates the total mass more accurately than the sector method but requires each frame to be considered individually, and is therefore more time-consuming.

3. Graduated cylindrical shell model (GCS) method. This method uses the forward modeling technique developed by \citet{thernisien2009} and \citet{thernisien2011} to define the outer boundary of the CME. The different parameters in the model apart from the radial distance are manually chosen from comparing the model shape with both STEREO COR-2 images simultaneously. These parameters are then fixed and the radial distance is varied over the sequence of images as the CME propagates. The outer edge which is traced out by the model is used in the same manner as the ROI method defined above. This new GCS method is used in our paper. 

\begin{figure}
\noindent\includegraphics[width=23pc]{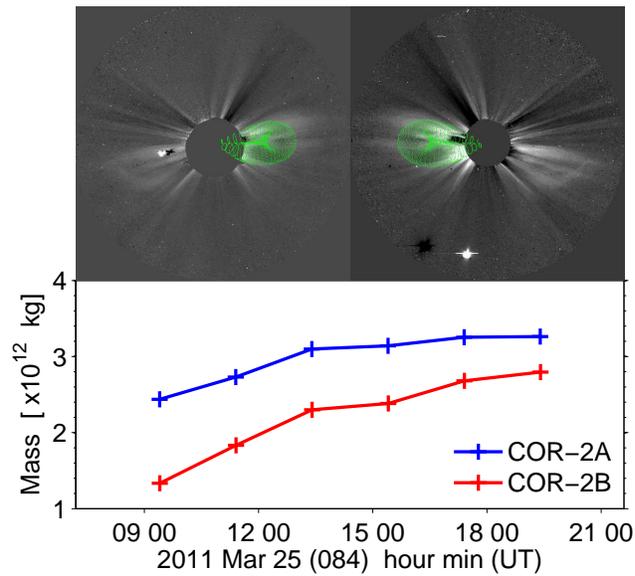}
\caption{(Top) Panels show the graduated cylindrical shell (GCS) model overlaid onto COR2A and B mass images at 15.24, 25 March 2011. (Bottom) Shows the measurements of the CME mass in COR2 by using the GCS model as the CME boundary.}
\label{cmeTotMass}
\end{figure}

The CME enters the COR-2A camera at 21.24 UT, March 24 2011 and approaches a mass of $\sim3 \times 10^{12}kg$ which is of the order of magnitude of a typical CME \citep{vourlidas2010}. The mass estimates as measured from the GCS model are displayed in figure \ref{cmeTotMass} and show the typical increasing-mass time profile due to the CME entering the field of view. This method enables the CME mass calculations to be made in a semi-automated fashion similar to the sector method, while using a more reliable ROI method for tracing the CME contour. In our example the mass of the CME was slightly overestimated by $\lesssim 5\%$ compared to using method 2. This is because we ensured that the entire CME structure was enclosed. For our event, which may not always be the case, it was relatively simple as the CME had well defined boundaries. However, as the CME propagated the leading edge became slightly flattened in comparison to the idealised GCS model. This meant the nose of the GCS was progressively further into the heliosphere than the observations. This meant a few extra pixel of mass were included but less than would have been encountered by method 1. It was found that the accuracy of the mass measurements compared to method 2 depended on ensuring that at least the entire area of the CME was included rather than minimising the surplus area of background solar wind. 

The GCS method used in this paper requires several parameters to be defined for each image. The results of propagation direction and flux rope axis were compared to the fixed-phi method (from remote observations) and constant-alpha force-free flux rope model (from in situ data; see \S ~\ref{caff2} for more details), respectively. The results were consistent with each other. The independently estimated propagation directions were within $10^{\circ}$ of each other and the three independently estimated flux rope axes were within $30^{\circ}$ 

Another option for measuring the mass flow is to investigate a fixed location in the heliosphere by defining a small narrow rectangular box a few pixels wide (slit method). This method emulates a time series of mass over a fixed location such as a spacecraft at the L1 position (see \S ~\ref{mtseries}). Currently, the minimal image processing used in this study does not allow for an accurate estimate to be made at L1. However, future studies using more advanced image processing and better tempo-spatial resolution as expected from Solar Orbiter should provide the necessary data to advance the techniques in preparation for a possible mission to L5.

	\subsection{Mass J-maps}\label{mjmap}
Originally developed for LASCO coronagraphic images, \citet{sheeley1999} calculated that a small plasma packet moving at uniform speed would have an apparent acceleration and deceleration which is dependant on 2 variables: the radial velocity, Vr, and the angle of propagation between the CME-Sun-spacecraft, $\beta$. Moreover, the observed acceleration profile (measured in time and elongation angle away from the Sun, $\alpha$) is unique; therefore by using an optimisation routine \citep[e.g.][]{savani2009, savani2010}, an estimate of the propagation direction of the CME can be made. This technique (called fixed-phi method) has been shown to be much more useful over a large range of elongation angles \citep{williams2009, davis2010} and produced very effective results in tracking CMEs from the Sun to planetary systems where they were detected in situ \citep[e.g.][]{davis2009, rouillard2009, mostl2009}. As the premise on which this technique is built relies on the idea of a spatially narrow plasma packet and not a large 3-dimensional (3D) object travelling through the heliosphere, other attempts have been developed to mitigate against some of the simplified assumptions \citep{kahler2007, lugaz2009, davies2012}.

\begin{figure}
\noindent\includegraphics[width=23pc]{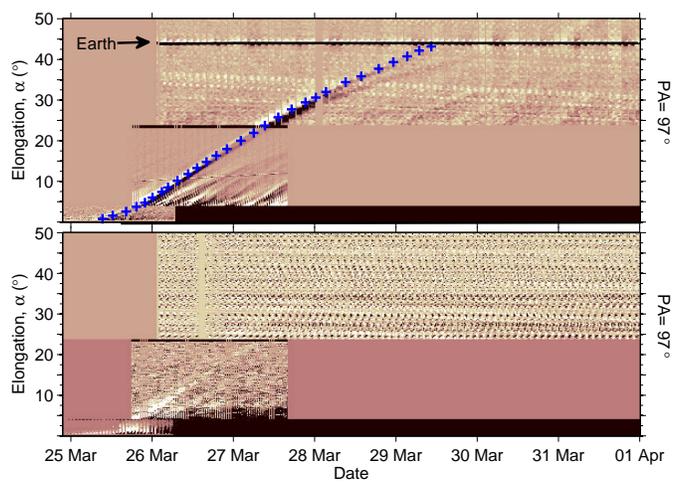}
\caption{J-maps created for the position angle of the Earth (PA=$97^{\circ}$). (Top) J-map displays the results from total brightness images with the location of Earth and the manually selected track of the CME's leading edge. (Bottom) Displays a mass J-map created from a sequence of mass images.}
\label{Jmapx2}
\end{figure}

The propagation direction is estimated by measuring the elongation angle as a function of time. This is most conveniently done by tracking a single feature within a CME through the field of view of all the cameras by using a time-elongation map (J-map). As a plasma packet can be assumed to propagate radially away from the Sun, the radial cuts used in J-maps can be varied to suite the CME direction if it is away from the ecliptic plane. In the case of our event we choose a P.A. of 97$^{\circ}$ from ST-A, which corresponds to the direction to Earth. Until this paper, these J-maps have been created by processing MSB images into a running difference sequence, as shown in the top panel of Figure \ref{Jmapx2}. The blue crosses are manually chosen from the image and are the points used in estimating the speed and propagation direction of the CME. Using the fixed-phi method the radial speed and propagation direction was estimated as 371 km/s and 80$^{\circ}$, respectively. The bottom panel displays the same time-elongation tracks but created with a sequence of mass images. As can be seen between the two versions, the white light J-map is clearer at identifying the propagation of the CME. This is partly because further smoothing and image processing was undertaken on the white-light J-map. It is hoped that further studies into mass J-maps will be used to provide a time series estimate of the mass propagating over the Earth in a process similar to the slit method described above. This is because the slit method would effectively represent a horizontal line along a mass J-map. Although the tracks in the mass J-map displayed in our paper is 'noisy', future studies may implement advanced image processing techniques that are currently under development \citep[e.g.][]{howardt2012} to find interesting discoveries.

The tracking of the mass estimates can be used with a minimal number of assumptions about the expansion process to estimate the density of the CME. With improvements to the image processing this information may be used as inputs to space weather forecasting models, instead of using L1 in situ data. Although this will clearly be less reliable than L1 data itself, it has the big advantage of being measured remotely and of the order of $\sim48$ hours in advance. This may prove to be a significant improvement for our forecasting capabilities.

\begin{figure}
\noindent\includegraphics[width=40pc]{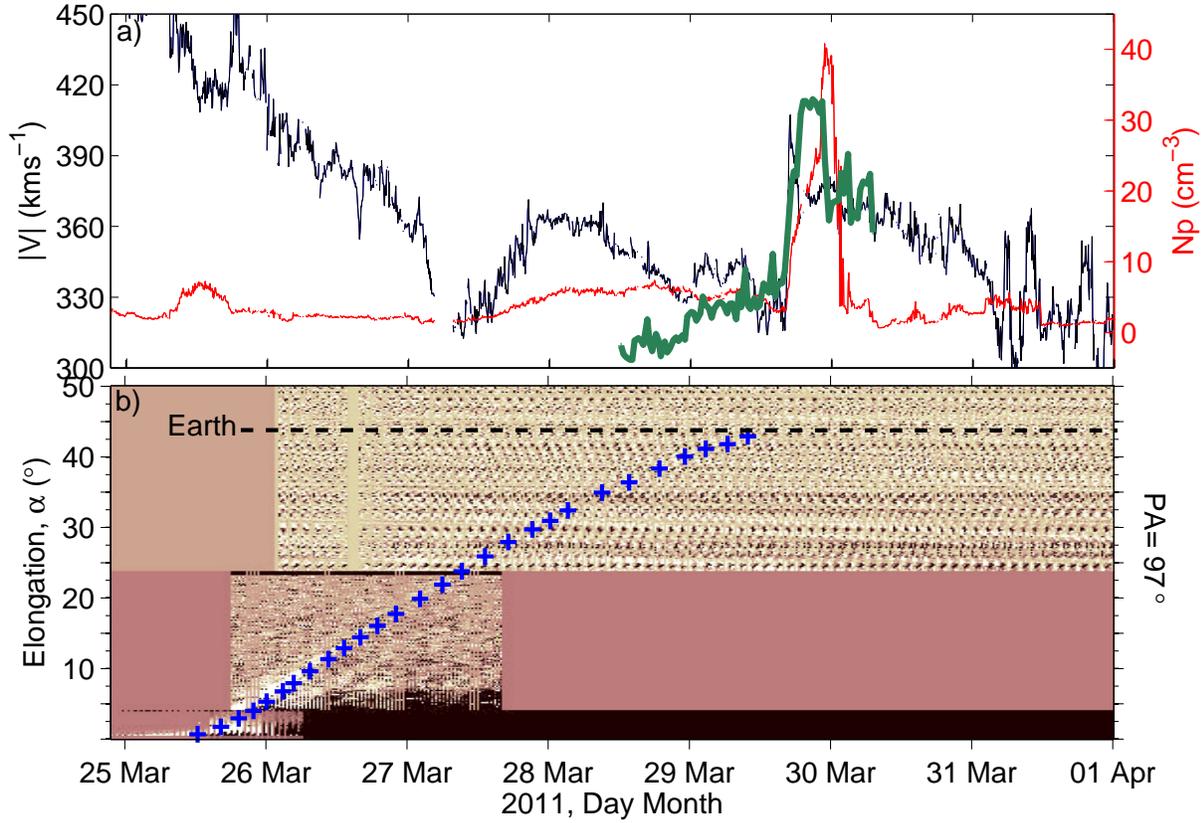}
\caption{(Top)The time series of the WIND proton density and velocity for the period 25 March to 01 April 2011. A normalised mass time series from the mass images  are overlaid onto the density axis (green). The mass time series is generated from the slit analysis at $3.8^{\circ}$ elongation within the COR2-A camera. The time series is ballistically propagated to L1 at a speed of 400km/s at the leading edge and 370km/s at the trailing edge, with a linearly decreasing speed profile. Qualitatively, the normailsed mass time series shows a remarkable resemblance to the in situ density of the CME sheath. (Bottom) The mass J-map with the manually selected CME leading edge.}
\label{rhoVelnJmap}
\end{figure}

Figure \ref{rhoVelnJmap} displays the mass J-map along with the in situ measurements of density and velocity at L1. The track, which was made from the white-light J-map, is overplotted onto the mass J-map along with a dashed line to show the position of Earth. The track clearly intercepts the position of Earth at the same time as the CME-driven shock (and the associated density increase from the sheath) arrives at L1. The shock arrival at L1 displays a sudden increase in momentum flux which then compresses the Earth's magnetosphere (see \S ~\ref{Earth}).

	\subsection{Mass time series}\label{mtseries}
\begin{figure}
\noindent\includegraphics[width=36pc]{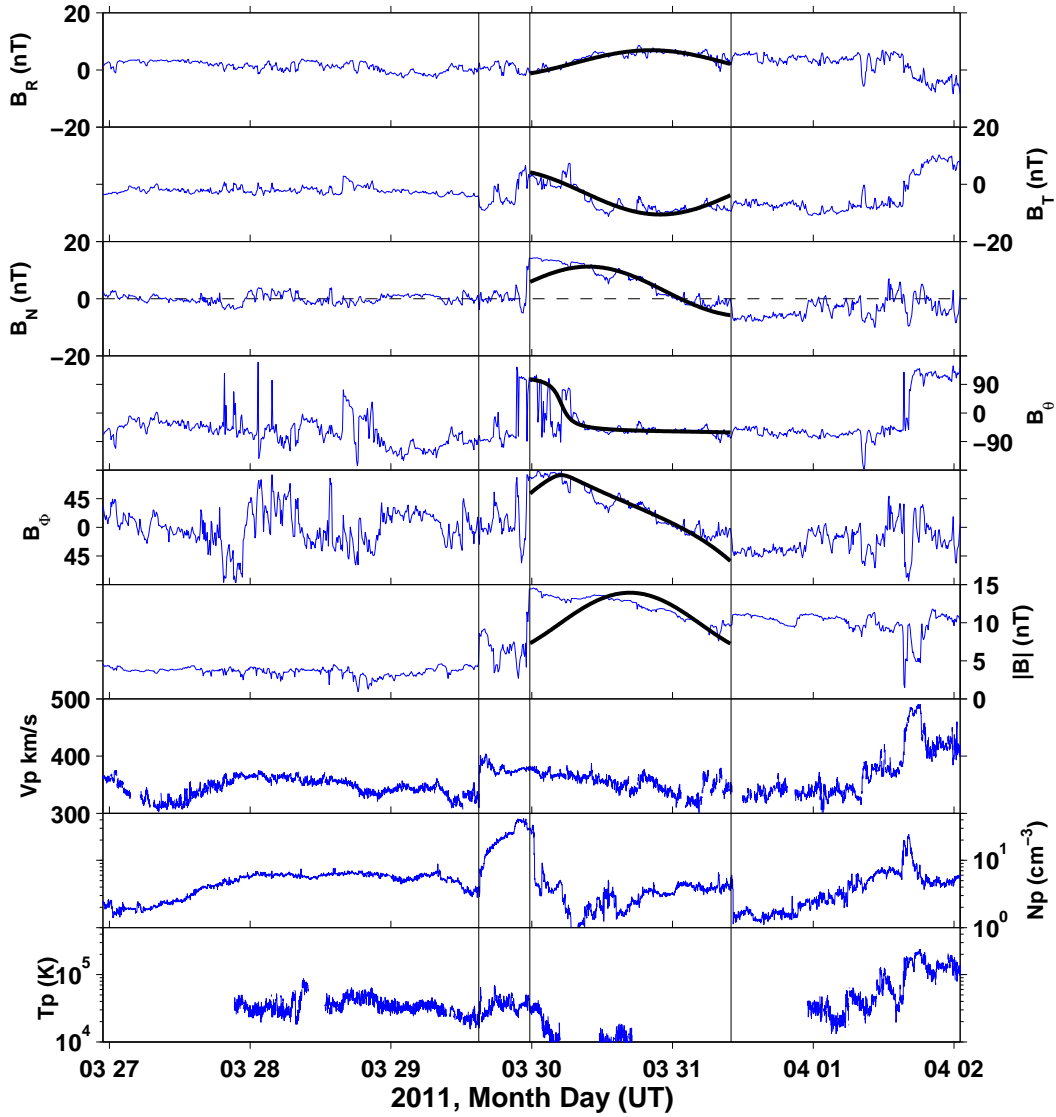}
\caption{In situ data from ACE and WIND at L1 point. The top 6 panels show the magnetic field in RTN cartesian and spherical coordinates system; followed by proton velocity,density and Temperature, respectively. The vertical lines from left to right display the positions of the CME shock, and magnetic flux rope leading and trailing edge.}
\label{inSitu}
\end{figure}

Under the premise that the CME momentum can be tracked and possibly be used as early solar wind input into space weather simulations, it is important to observe how a typical mass time series may look like and how it compares to the in situ density profile currently being used as simulation inputs. Figure \ref{rhoVelnJmap} shows a normalised times series of the mass measurements (green dashed curve) taken at $3.8^{\circ}$ elongation from ST-A, which is within the COR-2 field of view. The mass measurements were taken using the slit method at a plane of sky distance between 13.7-14Rs for each frame. The data was then ballistically time-shifted to the L1 position by assuming the CME travelled at 370km/s and was linearly expanding \citep{owens2005} so that the trailing edge of the mass measurements propagated 30km/s slower than the sheath leading edge (see also Figure \ref{inSitu} and \ref{Dst}). As our event initially propagates with a slow speed we are able to assume the CME was swept into the solar wind and merely advected out to 1 AU \citep{siscoe2006}. However for faster CMEs it is important to consider deceleration due to drag effects that change the arrival times at Earth \citep{gopalswamy2000, cargill2004}. The units of mass displayed have been normalised to suit the density profile. This has simply been carried out by dividing the measurements by $1.5\times 10^{11} g.cm^{3}$. A simple method of an expanding volume during the CME's propagation is used to justify this value (see Appendix \ref{vol} for more details).

Currently only minimally processed images are being used and even with our simplified propagation technique \citep[e.g]{owens2004,lugaz2012a}, the qualitative profile of the mass measurements made from remote observations is remarkably similar to that observed in situ. It is clear that CMEs may undergo interactions during propagation to Earth by either solar wind distortions \cite[e.g.][]{lugaz2008, savani2010}, deflections \citep{lugaz2012c, wood2012a} or possible rotations \citep[e.g.][]{Shiota2005, vourlidas2011, nieves2012}. This could be the cause of the larger predicted mass measurements within the CME when compared to the in situ values. Or more simply that the CME expanded more during the propagation and is more rarified by the time it reaches Earth.

\section{In situ observations}
The shock associated with our case study event was detected at L1 at 15.09 March 29 2011 with a jump in velocity of $\sim70$ km/s (the upstream speed was $\sim330$ km/s). Figure \ref{inSitu} shows the in situ paramaters of the solar wind during the propagating interplanetary CME. The three vertical lines indicate the locations of the shock front (14.58, 29th), CME leading edge (23.39, 29th) and rear edge (09.52, 31st) respectively. The CME leading and rear edge were manually chosen by focusing on looking for a duration that includes a smooth rotation in magnetic field and a discontinuity on density. The focus on the smooth field rotation is in order to produce reliable results from an in situ flux rope fitting process. The magnetic field vectors are displayed in the RTN coordinate system such that the Normal (N) component is the measure of the magnetic field in the out-of ecliptic direction (i.e. the North-South direction that is crucial in space weather predictions). This case study shows that the magnetic field in the CME and sheath region is predominately northwardly directed, and strongly so in the earliest half of the CME just behind the leading edge. This CME topology was specially chosen for our analysis as it allows our study to isolate the space weather effects (e.g. potential strength of GICs) that are caused by the momentum flux and not from the resulting magnetic reconnection from a southwardly directed interplanetary magnetic field (IMF).

	\subsection{Modelling results for March 2011} \label{caff2}
Figure \ref{inSitu} displays the magnetic field profile of the optimised CAFF model (see Appendix \ref{appFR} for details) within the top 6 panels as black curves. The top 3 represent the cartesian vectors in RTN coordinate system and the other 3 represent the field vectors in spherical coordinate system. The smooth rotation in the field is often more clearly seen in spherical coordinates while the importance of a southward Bz for predicting space weather events is better observed in cartesian coordinates. The orientation of the estimated flux rope axis direction is $(0.4,-0.7,0.6)$ in RTN and has a right handed chirality. The optimised parameters for the axial magnetic field and the impact parameter are 14.0 nT and 0.1, respectively. This indicates the spacecraft travel close-to but slightly above the FR axis. The mean square error between the optimised model and the data, $\chi$, was $ < 0.1$, which represents a good fit to the data \citep{lynch2003}. The model fits were also compared to a non-force-free elliptical model \citep{hidalgo2012} to show consistent results (e.g. the axis orientation were within $30^{\circ}$ of each other).

	\subsection{Dst index}
There is a variety of ground responses at Earth from space weather disturbances and measurements are often focused on different geographically localised processes/positions (e.g. due to local noon time, latitude or ground resistivity). The disturbance time (Dst) index measures the hourly values of the horizontal component of the Earth's magnetic field averaged from four near-equatorial geomagnetic observatories. The Dst index has historically been used as an approximation of the global response to a space weather disturbance. The fluctuations in the Dst closely relates to the ring current and the other current systems (including the magnetotail current) within the terrestrial environment. The inverse proportionality relationship between the horizontal component of the magnetic field and the energy content of the ring current is known as the Dessler-Parker-Sckopke relation \citep{dessler1959, sckopke1966}.

The stereotypical time profile of the Dst index during a geomagnetic storm driven by a fast CME displays an initial positive sharp rise (called the sudden storm commencement, SSC) which defines the arrival of the leading shock front onto the magnetosphere, a drop in value to zero (called the storm onset, SO), and then the main phase is characterised by a large negative decrease which represents the period of strong southward Bz magnetic field. The sheath region between the leading edge of a CME and the shock front is often considered to be related to the period between the SSC and the SO (initial phase) during a geomagnetic storm. However, the phases of a geomagnetic storm can be time shifted under conditions where a strong southward Bz persists within the sheath region.

\begin{figure}
\noindent\includegraphics[width=30pc]{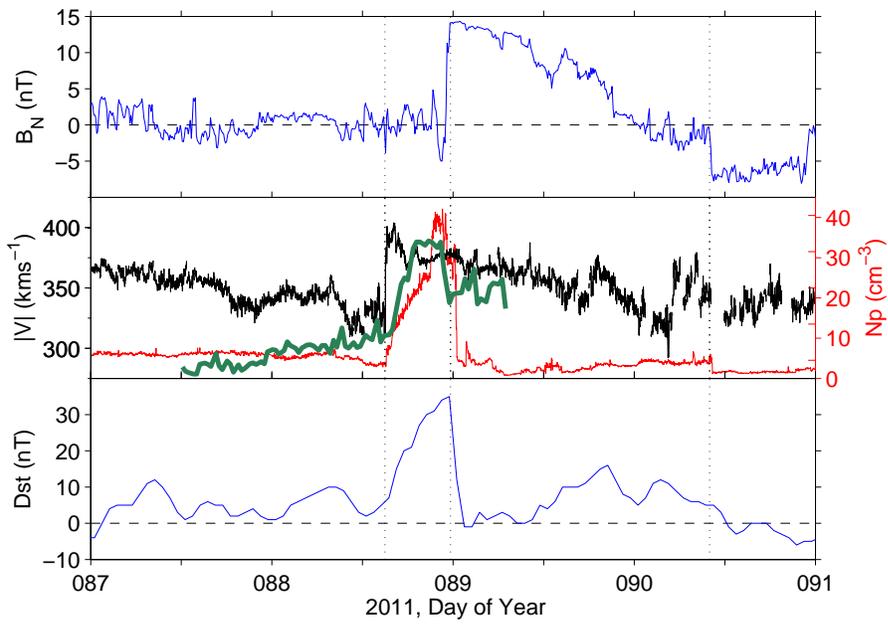}
\caption{A comparison between the in situ data from L1 (Bz component, velocity and density) to the averaged global response by the Dst index. The normalised mass time series from the COR2-A is displayed on the density axis as in figure \ref{rhoVelnJmap}. }
\label{Dst}
\end{figure}

The case study CME analysed in this paper was chosen to better understand the significance of its momentum flux as a driver of geomagnetic storms. As such, we chose to investigate a CME with a predominately northward Bz field. This allowed us to isolate the observed geomagnetic disturbance and assume the disturbance is solely due to the momentum flux and not due to reconnection between the CME and magnetosphere. For this reason, the usual main phase of a storm as seen in the Dst is not seen in our event (Figure \ref{Dst}). However, Figure \ref{Dst} shows a significant rise in the Dst value and sharp fall during the initial phase, which represents the location of a sudden increase in momentum flux from the CME sheath region. The sudden changes in the Dst shows that the momentum flux of the CME leading edge is capable of making sudden changes to the ground magnetic field (i.e. cause a large dB/dt).

\section{Terrestrial response}\label{Earth}
Sophisticated MHD simulations in 3D are becoming an increasingly effective tool for modelling solar wind transients such as CMEs and for predicting their geo-effectiveness at Earth. In this paper we have employed the Space Weather Modeling Framework \citep[SWMF]{toth2005} package which was executed at the Community Coordinated Modeling Center (CCMC) and operated at NASA Goddard Space Flight Center. The solar wind input data was chosen from the WIND spacecraft at L1 and was ballistically mapped to the outer boundary of the BATSRUS magnetospheric MHD model that was coupled to the Rice Convection Model (RCM) in our simulations. The auroral conductances are driven by solar irradiance observations of F10.7 and field-aligned electric currents.

\begin{figure}
\noindent\includegraphics[width=23pc]{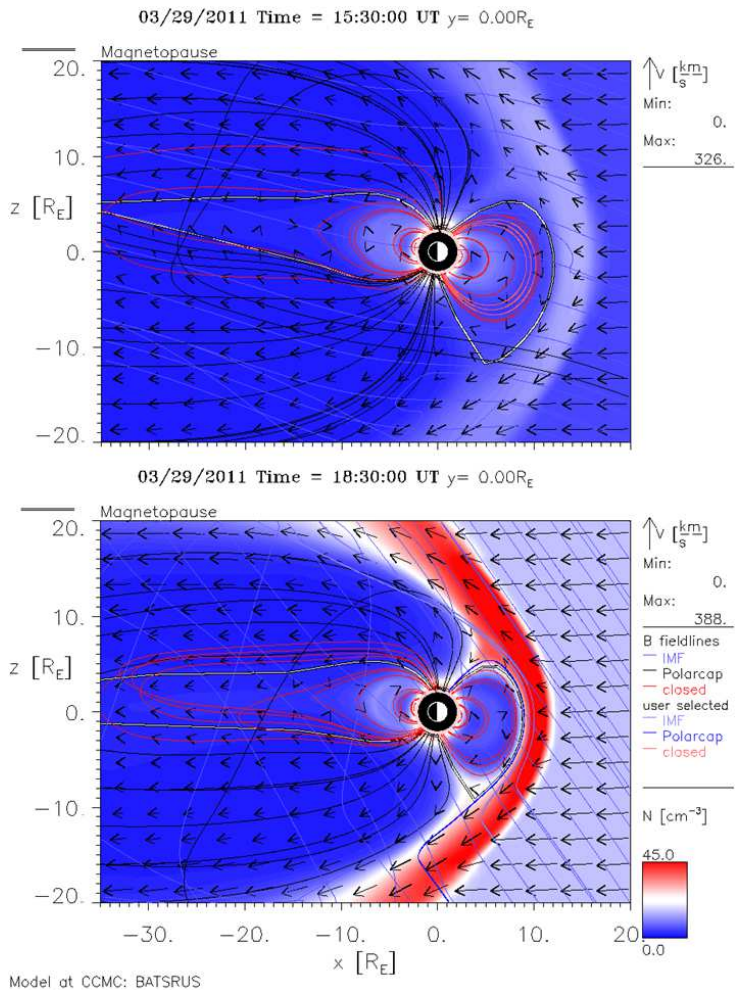}
\caption{2D cut of the Earth's magnetosphere and bow shock from the BAT-R-US simulation at CCMC. The colour scale of the density is the same in both panels. (Top)Displays the shape and structure of the magnetosphere prior to the arrival of the CME shock. (Bottom) Shows the compressed magnetosphere at the time when the CME sheath is affecting the bow shock.}
\label{ccmc}
\end{figure}

Figure \ref{ccmc} displays a 2D cut of Earth's magnetosphere (a) before the arrival of the CME-associated shock front and (b) when the sheath region of the CME is travelling over Earth's bow shock. The colour table represents the density of the plasma and a few selected magnetic field lines are drawn to help distinguish between the terrestrial and heliospheric systems. The vectors show the solar wind direction. In the 3 hours between the frames shown in Figure \ref{ccmc} the bow shock is severely compressed from a location of approximately 16Rs to $\sim10$Rs. This compression significantly increased the density within the magnetosheath and was due to the sudden arrival of a larger momentum flux from within the sheath region of the CME.

Currently this study uses the in situ data measured at L1 as the inputs into the SWMF simulations in order to predict realistic geomagnetic disturbances. However it is envisaged that with further development to the mass images, a reasonable density time-series calculated remotely may be used as an input into the BATSRUS code (see \S ~\ref{mtseries}). This may prove to be a valuable tool in improving any early warning systems by being able to provide an observationally-predicted result that is at least $\sim24$ hours earlier than is currently possible. 
% 	\subsection{Space weather simulations}

	\subsection{Observations of magnetic fluctuations}
In order to better understand the geomagnetic effects driven by CMEs and in particular to predict the socio-economic impacts from GICs it is important to study the localised effects observed at specific locations on Earth during the arrival of the CME. In this paper we study the momentum flux of CMEs and therefore choose to investigate the equatorially based ground stations. The higher latitude stations are likely to display magnetic field fluctuations that can be partly associated with auroral magnetosphere-ionosphere dynamics, and are therefore not the focus of this paper. The shock arrival at L1 ($\sim15.00$ UT) defines the start of the geomagnetic disturbance. We investigate two locations on Earth that are determined to be locally noon and midnight at the shock arrival. The Vassouras (VSS) station in Brazil is used as the locally noon station. %, and is one of the four stations used in averaging the Dst index. 
The Kanoya (KNY) and Kakadu (KDU) stations in Japan and Australia, respectively, are studied as the locally midnight stations for our investigation. 

\begin{figure}
\noindent\includegraphics[width=35pc]{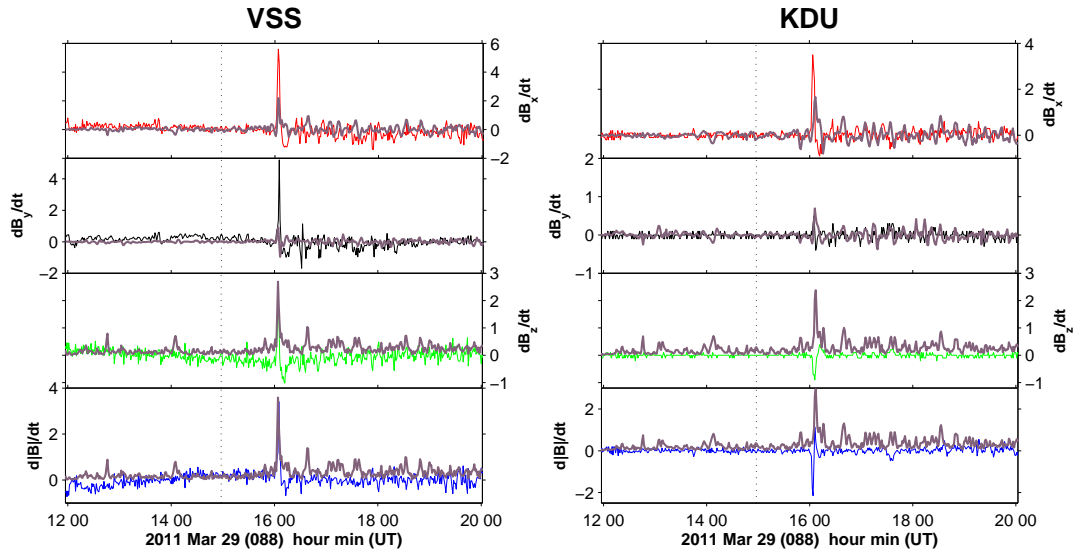}
\caption{The time-derivative of the magnetic field measured at the low-latitude Vassouras (Brazil) and Kakadu (Australia) ground stations. These stations represent the local noon and midnight of the CME shock front onto the bow shock respectively. The purple curves overlaid onto the data are the estimated values at these locations from the CCMC simulations.}
\label{grndStn}
\end{figure}

In order to investigate the possible effects that a CME may have on GICs, the time-derivative of the ground magnetic field (dB/dt) must be studied. Figure \ref{grndStn} displays the dB/dt at both local times and shows significant fluctuations at the arrival of the shock onto Earth's system. The vertical dashed line displays the time the shock arrived at L1 and therefore the delay of $\sim1$ hour on the ground stations predominately represents the propagation time of the solar wind between L1 and the bow shock. The data is displayed with one minute temporal resolution in the cartesian geographic (GEO) coordinate system provided by INTERMAGNET (www.intermagnet.org).

	\subsection{Simulated dB/dt}
Quantifying and predicting ground magnetic field perturbations are vital to the space weather community. As such, the CCMC has developed a tool that is able to extract the ground magnetic field perturbations from the global MHD model outputs by integrating the results from the magnetospheric and ionospheric current systems. In particular, a summation of four separate current systems is used to make the predictions: 1. the current system in the magnetosphere (and magnetotail) above 2.5 Re; 2. field aligned currents between 2.5 Re and 110 km; 3. Hall current from the ionosphere; 4. Pederson currents in the ionosphere \citep{rastaetter2004,pulkkinen2010b}. The performance and metric-based analyses between various modelling approaches were part of the Geospace Environment Modeling (GEM) $2008-2009$ challenge and is reported by \citet{pulkkinen2011}.

Figure \ref{grndStn} shows the predicted estimates of the ground field as purple curves over-plotted on the data. The simulated time series have been shifted earlier in time by 15 minutes which we attribute to a small uncertainty in the input data which resulted from ballistically shifting the input L1 data to the outer boundary of the simulation. We show that the simulated data also displays a significant spike in the dB/dt at the storm commencement which coincides with a similar magnitude to the `ground truth' observed by the magnetometer ground stations. Qualitatively, the biggest limitation in the simulated results appears to occur in the inaccuracy of the negative dBz/dt component seen during the time derivative `spike' This could be due to the Bz component being especially sensitive to geomagnetic induction effects which are not taken into account in our simulation.

It is worth noting that only mid and high latitude magnetometer stations were included in the earlier GEM challenges even though the Dst index addresses the low-latitude disturbances. This is due to the global MHD approach being implemented in the first-principle models; they could only capture the ionospheric output at high latitudes by using the Biot-Savart law to integrate over the ionospheric electric currents system. In this paper we have alleviated such a constraint by coupling the global MHD models to the inner magnetospheric models and thereby capturing the ring current dynamics and magnetospheric current systems; therefore providing the required ionospheric response at low latitudes \citep{yu2010}.

	\subsection{Geomagnetically induced currents, GICs}
Reliable estimates of GICs and the geoelectric field requires accurate knowledge of the local geological conditions as well as the dB/dt. As the global distribution of the conductivity from the surface to the upper mantle (depths of several hundred kilometers) is not well known, estimating GICs can not be arbitrarily made for any location on Earth. However, for local environments that have historically been susceptible to GIC events \citep [e.g.][]{pulkkinen2005, ngwira2008, pulkkinen2010, torta2012}, the ground structure is known and therefore this study could be replicated for larger geo-effective CME events and early warning predictions can be made as to their level of susceptibility. It is envisaged that as the ground structure for more locations around the globe become recognized, the framework presented in this paper will provide the necessary steps to improve GIC forecasting. Of course, a simple approach may be created to estimate the extrema of possible GICs by using realistic extreme ends of the conducting (British Columbia, Canada) and resistive (Quebec, Canada) ground structure \citep{pulkkinen2008}. But detailed investigation into these possibilities goes beyond the scope of this work.

\section{Discussion}\label{Disc}

Currently observations of the solar magnetic fields are used as inputs for the background solar wind when simulating the heliosphere for space weather predictions. However the CME itself is usually set to an approximate and generic size. Details of the CME that is included in the magnetospheric simulations are only provided by measurements made in situ at L1, which is $\sim1$ hour before its arrival at the Earth's bow shock. By using remote observations from coronagraphs ($\sim15 Rs$), observational estimates of the CME can be made for both the Enlil heliospheric simulation and as early initial-attempt inputs to BATSRUS magnetospheric simulations. The use of remote observations means that BATRUS simulations can be carried out $\sim48$ hours prior to waiting for the CME to propagate to L1.

The focus of this paper was to estimate the CME mass as a time series close to the Sun; which was then ballistically propagated to L1. While previous studies have been carried out to better estimate the arrival times of a CME \citep{owens2004, taktakishvili2009, taktakishvili2011} and extensive plane-of-sky speed measurements have been made over a large number of CMEs \citep{stcyr2000, yashiro2004}. These studies could be used to remotely estimate a velocity time series which can then be used to propagate the CME to L1. The ability to remotely estimate both the velocity and mass allows early prediction of the momentum flux arriving to Earth. 

Figure \ref{rhoVelnJmap} and \ref{Dst} display the remotely-observed mass time-series which have been artificially normalised to suit the in situ number density. The mass time-series were arbitrarily divided by $1.5 \times 10 ^{11} g$ in order to display a qualitative profile of the same order of magnitude. Clearly, if these mass time-series are to be used as inputs to early space weather simulations then an appropriate method to scale the mass estimates to a density value is required. Three basic methods can be used: 1. Defining the CME's 3D volume by, for example, by using the GCS model \citep{thernisien2006}. This volume can then be radially propagated out to 1 AU where the density can be estimated. 2. Using the empirical formulas proposed by \citet{vourlidas2010}. 3. Using a combination of white light images and off-limb spectroscopy to directly measure the densities across the CME body. Future investigations may attempt to solve the most appropriate method with more events. Method 1 has been used in Appendix \ref{vol} to justify the order of magnitude used in this paper. With appropriate density and velocity estimates made remotely to first approximation, the results can be adjusted in an ensemble method for making space weather predictions along with variety of magnetic field estimates \citep [e.g.][]{lin2000, gopalswamy2011, Savani2012b}.

\section{Conclusions}

In this paper we show the first results for producing density measurements from remote observations that are comparable to their equivalent in situ time-series. Therefore this paper for the first time shows that a time series of data can be estimated remotely and be used to make forecasts of GICs at Earth. While a previous attempt by \citet{pulkkinen2009} to forecast GICs from remote observations only used a generic pressure pulse for a CME within the Enlil model. We also confirm that the compression of the magnetosphere from the momentum flux of a CME is a significant variable in predicting geomagnetically induced currents. Our results are capable of producing qualitatively reliable estimates of the densities upstream of Earth's bow shock. These results indicate the possibility of using remote observations at a heliocentric distance of $\sim15 Rs$ to estimate the solar wind density profile arriving at Earth and show that these estimates can be used as part of a preliminary early warning system for space weather predictions.

This paper focuses on a case study event of a CME which displays a strong northward Bz component in the magnetic field rather than a fast geo-effective event. The CME propagated in the inner heliopshere between 25-31 March 2011 and is observed in both ST-A and ST-B. The event allowed our study to attribute all the ground based effects at low latitudes to the compression of the magnetosphere and not to magnetic reconnection in the case of a southward directed Bz. The compression of the magnetosphere is found to be the result of the larger momentum flux (larger density and velocity) in the sheath region of the CME.

Previous studies have estimated the mass of a CME as it travelled through the field of view by either manually measuring the region of interest (ROI) around the CME for each individual frame or by estimating it within a fixed sector and thereby assuming the inclusion of the additional solar wind mass is not significant in relation to the CME mass. Here, we improve the process by semi automating the ROI while refraining from the manual selections within each frame. We do this by using the graduated cylindrical shell model \citep{thernisien2011} to define the ROI and manually fixing the parameters by eye at the beginning and then by only varying the radial distance for each consecutive frame.

To track the CME mass from the remote observations to L1 we employed the J-map technique for a sequence of mass images in the STEREO data. The J-map technique has previously been extensively used with white light images to estimate the propagation direction and arrival time of a CME \citep[e.g.][]{sheeley1999, rouillard2009}. However, we show that the mass J-map technique can be used to highlight the mass intensities travelling towards Earth. As mass images are estimated from base-difference frames rather than running differences, the background star field becomes significant in the HI-2 field of view. In this paper we have chosen to use a minimal amount of image processing in order to estimate a base confidence level for the uncertainties in the mass values. Future studies may consider a comparison between the minimal processed images to more advanced techniques.

As the mass images in the vicinity of Earth (HI-2 field of view) was not clearly visible in the images due to the interference of the background star field, we chose to estimate the mass as a time series at a fixed heliocentric location within the COR-2A field of view (plane of sky distance of 14 Rs). The mass time series was then time shifted by assuming a leading edge speed of 400km/s and having a linear expansion profile such that the trailing edge of the mass profile was travelling at 370km/s. The qualitative comparison of our estimated mass profile to the in situ density measurements at L1 are remarkably similar for the sheath region between the shock front and leading edge. Future studies to investigate CMEs may not necessarily have the capabilities to track the CME along the entire Sun-Earth line. As such, our ballistic propagation approach for the mass measurements from $\sim15Rs$ to L1 may serve as a useful tool for estimating a time series for the mass.

%The inclusion of remotely observed density estimates into the NASA space weather modeling framework (SWMF) can provide a rough order-of-magnitude estimate more than 48 hours in advance to the current process which waits for the in situ measurements.

%HI observations for CMEs in the future after the STEREO era \cite{lugaz2012b}

%% ------------------------------------------------------------------------ %%
%
%  APPENDIX
%
%% ------------------------------------------------------------------------ %%
% \appendix resets counters and redefines section heads
% but doesn't print anything.
\appendix

\section{Density Normalisation}\label{vol}
In this paper we have chosen to artificially divide the time series of the mass measurements by a constant value of $1.5\times 10^{11} g$. This was used to convert a mass times series that was estimated remotely into a density time series which was later compared with the in situ measurements. Here we carry out some preliminary work to justify the number used with a simple volume expansion method. The work below is intended to provide an order of magnitude justification, however further work would benefit from a more detailed approach as suggested in \S ~\ref{Disc}.

\begin{figure}
\noindent\includegraphics[width=35pc]{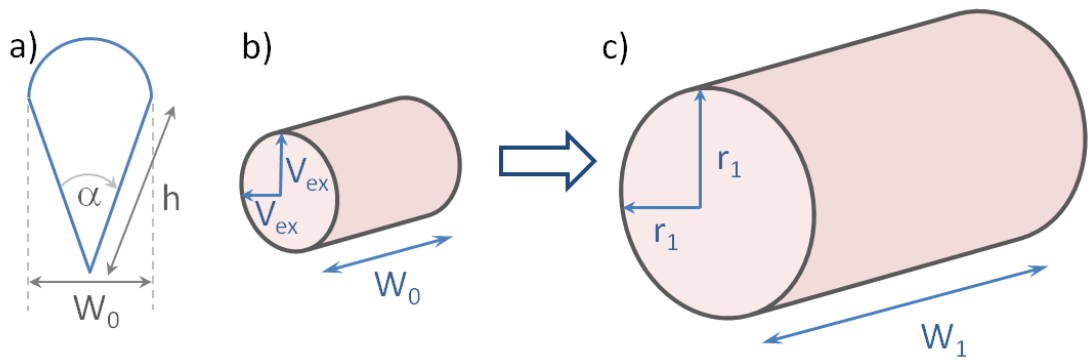}
\caption{a) Shows the parameters from \citet{thernisien2006} that are used to estimate the length of the CME along the line of sight. b) Is the initial volume element defined to be observed with the COR 2A camera. c) Shows the final dimensions of the volume element at L1, where the in situ measurements are taken.}
\label{figA1}
\end{figure}

We assume that a small volume (defined by the size of a few pixels in the COR 2A camera) measured at $\sim10 Rs$ has a shape of a cylinder, such that the circular cross section is within the plane of sky and the length ($W_0$) is defined along the line of sight. First we assume that the length of the cylinder can be estimated from the GCS model (see Figure \ref{figA1} for details). At 15.24 UT on March 25 2011, the height of the legs (h) and the half angular width ($\alpha$) from GCS as defined by \citet{thernisien2006} was estimated as 9.78 Rs and $32.5^{\circ}$, respectively. With trigonometry, we calculate that the cylindrical length should be 10.51 Rs. For this time, we noted that the plane of sky position of the CME nose was 10.5 Rs. The same process was carried out for an image at 17.24 UT to deduce that the heliocentric distance of the CME was to first approximation equal to the cylindrical length, $W_0$. Therefore, at L1 where the in situ measurements are made, we estimate the cylindrical length as $W_1=210 Rs$.

To estimate the cross sectional area we assume the cross-section expands at a uniform speed of $V_{ex}= 30 km/s$ (which we estimated from the in situ measurements). The total time of expansion, t, is the same as the propagation time and dependant on the bulk flow speed which we assume to be 370 km/s. Assuming the propagation distance, R =200Rs, we find that the radius of the circular cross section at L1 is 16.22Rs and follows
\begin{equation}
r_{1}= \left(  \frac {V_{ex}}  {V_{bulk}}  \right) \times R  \\ .
\end{equation}

Therefore our final volume element ($\Gamma_{f}$) can be estimated by, 
\begin{equation}
\Gamma_{f}= \pi  \left(  \frac{V_{ex}}{V_{bulk}}  \right)^2 	R^2 W_1  \\
\end{equation}
to give $5.85 \times 10^{37} cm^{3} $.

It therefore follows that a first approximation for a normalisation constant ($\kappa _0$) is,
\begin{equation}
\kappa _{0}=  \frac{1}{m_p \Gamma_{f} } \\ .
\end{equation}

From the constants used in our example, $\kappa _{0}= 1.02 \times 10^{-11} g^{-1}cm^{-3} $. This is similar to the normalisation value of $1.5 \times 10^{11}$ used in this paper and is certainly of the correct order of magnitude. This small discrepancy might be solved with more advanced calculations.

\section{Flux rope fitting}\label{appFR}
% Appendix A: Here Is Appendix Title

Simple determination of parameters from a model fitting procedure is one of the best ways to quickly estimate the global properties of an interplanetary CME from in situ measurements. The first model to be successfully optimised to MCs was a constant $\alpha$, force-free (CAFF) flux rope model by \citet{lepping1990}. Since then, several other attempts have been made to improve the results between observations and models \citep{owens2006, mulligan2001, hidalgo2002, hidalgo2012, owens2012}. We use a simplified modification of the \citet{lepping1990} model. The magnetic field of the model is described by the cartesian form of the Bessel function for a force-free flux rope. In fitting the flux rope model to the data, the first step is to estimate the orientation of the cylindrical axis. Then by determining the chirality and varying both the distance of closest approach and axial magnetic field strength as free parameters within a computational code, we are able to make predictions of the ICME parameters.

The method employed to determine the MC axis direction involves a technique called minimum variance analysis, MVA \citep{sonnerup1967, steed2008}. MVA was originally developed to determine the local normal for a tangential discontinuity (TD) in the magnetic field. This method calculates the orthogonal set of vectors in which the variance of the magnetic field in question is at a minimum (e1), maximum (e3) and intermediate (e2). For each of these eigenvectors there is a corresponding eigenvalue, often represented as $ \lambda _{1}, \lambda _{3}$ and $\lambda _{2}$ respectively.

For the purposes of investigating TDs the minimum eigenvector calculated provides the normal to the plane. Here, the ratios of eigenvalues ($\lambda _{1}$) and the other two are of concern as they indicate the reliability in the normal. For investigating the axial direction of a MC, we must concern ourselves with the intermediate direction, with MVA applied to the time period that contains the flux rope only. The axis of the MC therefore lies in the intermediate direction, but the distinction of the field variance along the coordinate axes reduce as the spacecraft trajectory moves further from the central MC axis.

When investigating the intermediate direction, the eigenvalue ratios between $\lambda _{2}/\lambda _{1}$ and $\lambda _{3}/\lambda _{2}$ are of concern. Ideally both ratios should be $>10$, but recent work using MVA have shown ratios of greater than 2 are adequate in the ICME context \citep{bothmer1998}. If the ratio of either $\lambda _{2}/\lambda _{1}$ or $\lambda _{3}/\lambda _{2}$ is small then the axial direction becomes ambiguous within the plane of the two eigenvectors.

The scenario above is described for the ideal case, where the spacecraft passes through the centre of the flux rope. However, this is usually not true. The distance of closest approach, $Y_{0}$, is an important parameter in determining the predicted field profile. It is defined in dimensionless units as the closest approach distance divided by the radius of the flux rope. Thus $Y_{0}$ varies between 1 and -1 for a spacecraft travelling above and below the MC centre, respectively. Rees [2003] simulated idealised flux ropes with spacecraft trajectories away from the centre and then carried out MVA analysis over the time profiles. He found that the error in the axis angle increased with an increasing $|Y_{0}|$. The results suggest that an impact parameter of $>0.5$ should be dealt with caution as the error may be of the order of $\sim 15^{\circ}$. The uncertainty in ascertaining the axis orientation \citep{alhaddad2012} is one of the reasons why more complex models choose to begin their analysis with MVA and then later find the optimal direction by introducing 2 extra free variables.

High frequency noise is a source of error for MVA in the interplanetary CME context. MVA analyses the variance in the magnetic field, therefore large noise fluctuations can yield incorrect eigenvectors. To limit this effect, 10 or 15 minute averaged data are often used when fitting a modelled rope to the data. It has been shown for ‘good’ events the time resolution between 10 and 60 minutes has a negligible effect on the axis orientation \citep{lepping2003}. Subtly, the lower resolutions are often used for large events, with a reason to maintain the total number of data points within the MC to about N=40 \citep{lepping2006}. Although this is not essential, it is desirable when comparing various cases against each other in a statistical manner.

Once the axis of the model MC is determined, the in situ magnetic field data are rotated into the frame of the MVA axes. A sequence of flux ropes are then created and compared to the data by analysing the mean square error ($\chi$) coefficient. The $\chi$ coefficient measures how well each rope fits to the data and is defined in the same manner as \citet{lynch2003}.

The time interval between the start and end of the MC is identified manually by eye. This defines the data range which is compared to the model and fixes the size of the MC model. By using the distance of closest approach, the axial field strength and chirality (a concept of rotation in the magnetic field, either clockwise or anticlockwise) as free parameters, various simulated MCs can be generated. Each simulated MC is compared to the data by measuring the $\chi$ coefficient. The simulated MC with the smallest $\chi$ value is regarded as the best fit result. The free variables that create the model are varied by using a downhill simplex method developed by \cite{nelder1965}; this is a non-linear optimisation routine designed to minimise the $\chi$ coefficients. This approach is faster than a standard grid search and can be easily manipulated to determine the free variables to larger significant figure. The optimised free parameters are the outputs of the model. These results are used again to re-create the optimised model. The field vectors from this model are then rotated into the spacecraft frame and plotted on top of the observed data for visual confirmation.

It is worth noting here that the CAFF model described above assumes a static ICME. That is to say, the model plotted shows a time series obtained by taking a radial cut through a flux rope at a fixed time. Many ICMEs observed at terrestrial distances take $\sim 24$ hours to transit over a spacecraft, therefore it is important to note that the ICME ‘seen’ at the end of the transit has evolved and is different from the same object at the beginning of the transit. A more accurate representation would be to include a time series of a flux rope evolving in time past a fixed point in space.

%% ------------------------------------------------------------------------ %%
%
%  ACKNOWLEDGMENTS
%
%% ------------------------------------------------------------------------ %%
\begin{acknowledgments}
NPS thanks Neil Sheeley and  Adam Szabo for their collaborative assistance. This research was also supported by the NASA Living With a Star Jack Eddy Postdoctoral Fellowship Program, administered by the UCAR Visiting Scientist Programs and hosted by the Naval Research Laboratory. AV is supported by NASA contract S-136361-Y. The global MHD simulations used in this work were carried out at the Community Coordinated Modeling Center (CCMC) operated at NASA Goddard Space Flight Center. The authors wish to acknowledge Lutz Rastaetter, David Berrios and the rest of the CCMC staff for their generous support throughout the work discussed in the paper. The results presented in this paper rely on the data collected at Vassouras, Kakadu and Kanoya. We thank Observatorio Nacional, Geoscience Australia and Japan Meterological Agency, for supporting its operation and INTERMAGNET for promoting high standards of magnetic observatory practice (www.intermagnet.org).

\end{acknowledgments}

\end{article}

%% Enter Figures and Tables here:

% When submitting articles through the GEMS system:
% COMMENT OUT ANY COMMANDS THAT INCLUDE GRAPHICS.
% DO NOT USE \psfrag or \subfigure commands.
% Figure captions go below the figure.
% Table titles go above tables; all other caption information
%  should be placed in footnotes below the table.

% DRAFT figure/table, including eps graphics

%% ------------------------------------------------------------------------ %%
%
%  FIGURES
%
%% ------------------------------------------------------------------------ %%

% \end{document}
%
% \begin{table}
% \caption{}
% \end{table}
%
% ---------------
% TWO-COLUMN figure/table
%
% \begin{figure*}
% \noindent\includegraphics[width=39pc]{samplefigure.eps}
% \caption{Caption text here}
% \end{figure*}
%
% \begin{table*}
% \caption{Caption text here}
% \end{table*}
%
% ---------------
% EXAMPLE TABLE
%
%\begin{table}
%\caption{Time of the Transition Between Phase 1 and Phase 2\tablenotemark{a}}
%\centering
%\begin{tabular}{l c}
%\hline
% Run  & Time (min)  \\
%\hline
%  $l1$  & 260   \\
%  $l2$  & 300   \\
%  $l3$  & 340   \\
%  $h1$  & 270   \\
%  $h2$  & 250   \\
%  $h3$  & 380   \\
%  $r1$  & 370   \\
%  $r2$  & 390   \\
%\hline
%\end{tabular}
%\tablenotetext{a}{Footnote text here.}
%\end{table}

% See below for how to make landscape/sideways figures or tables.

\end{document}